\documentclass[aps,onecolumn,showpacs,showkeys,nofootinbib]{revtex4}
\usepackage{epsfig}
\usepackage{amsmath}
\usepackage{amsfonts}
\usepackage{amssymb}
\usepackage{graphicx}
\usepackage{colordvi}
\begin{document}
\begin{flushleft}
\end{flushleft}

\title{Photon frequency shift  in curvature based Extended Theories of Gravity}

\author{S. Capozziello$^{1,2,3,4}$\footnote{ e-mail address: capozziello@na.infn.it (corresponding author)}, G. Lambiase$^{5,6}$\footnote{e-mail address: lambiase@sa.infn.it}, A. Stabile $^{2}$\footnote{e-mail address: arturo.stabile@gmail.com}
  An. Stabile$^{5,6}$\footnote{e-mail address: anstabile@gmail.com}}

\affiliation{$^1$Dipartimento di Fisica, Universit\`{a} di Napoli "Federico II", Complesso Universitario di Monte Sant'Angelo, Edificio G, Via Cinthia, I-80126, Napoli, Italy}
\affiliation{$^2$INFN Sezione di Napoli, Complesso Universitario  di Monte Sant'Angelo, Edificio G, Via Cinthia, I-80126, Napoli, Italy}
\affiliation{$^3$Scuola Superiore Meridionale, Largo S. Marcelllino 10, I-80138, Napoli, Italy.}
\affiliation{$^4$Tomsk State Pedagogical University, ul. Kievskaya, 60, 634061 Tomsk, Russia.}
 \affiliation{$^5$Dipartimento di Fisica ``E.R. Caianiello'',
  Universit\`{a} degli Studi di Salerno, via G. Paolo II, Stecca 9, I
  - 84084 Fisciano, Italy}
  \affiliation{$^6$Istituto Nazionale di
  Fisica Nucleare (INFN) Sezione di Napoli, Gruppo collegato di
  Salerno}

\begin{abstract}

We study the frequency shift of photons  generated by rotating gravitational sources in the framework of curvature based Extended Theories of Gravity.  The discussion is developed  considering  the weak-field approximation. Following a perturbative approach, we analyze  the process of exchanging photons between Earth and a given satellite, and we find a general relation to constrain the free parameters of gravitational theories. Finally, we suggest the Moon as a possible laboratory to test theories of gravity by future experiments which can be, in principle, based also on other Solar System bodies.

\end{abstract}
\date{\today}
\maketitle

\section{Introduction}

Several observational data probe that the Universe appears spatially flat and undergoing a period of accelerated expansion \cite{riess, ast, clo, spe,carrol,sahini}. This picture is dynamically addressed if
two unrevealed ingredients are considered in order to achieve the agreement with observations: the {\sl dark energy} at cosmological scales and the {\sl dark matter} at galactic and extragalactic scales. The first is related to the accelerated expansion, the second to the clustering of structure. However, no fundamental particle, up today, has been clearly observed in view   to explain these ingredients despite of the huge experimental and theoretical efforts to address phenomenology. An alternative viewpoint is considering the possibility to explain large scale structure and accelerated expansion as gravitational phenomena.

Recently,  without introducing any exotic matter, Extended Theories of Gravity (ETG) \cite{Felix} have been considered as an alternative approach to explain  the galactic rotation curves and the cosmic acceleration \cite{Nojiri,vasilis, felice}. The approach  results from  effective theories aimed to deal with quantum fields in curved space-time at ultraviolet scales which  give rise to  additional contributions with respect to General Relativity (GR)  also at infrared scales: in this perspective,   galactic, extra-galactic and  cosmological scales can be affected by these gravitational corrections without requiring  large amounts of unknown material dark components. In the framework of ETG,  one may consider that the gravitational interaction acts differently at different scales, while the results of GR at Solar System scales are preserved. In other words, GR is a particular case of a more extended class of theories. From a conceptual viewpoint, there is no  \emph{a priori} reason to restrict the gravitational Lagrangian to a linear function of the Ricci scalar minimally coupled to matter~\cite{mag-fer-fra}.

However, ETG are not only curvature based but  can involve also other formulations like  affine connections independent of the metric, as in the case of  metric-affine gravity  \cite{ext1} or   purely affine gravity \cite{ext2}.  Metric-affine theories are also the Poincar\'e gauge gravity \cite{ext3}, the  teleparallel gravity based on the Weitzenb\"ock connection \cite{ext4,ext5}, the  symmetric teleparallel gravity \cite{ext6}. In summary, the debate on the identification of variables describing the gravitational field is  open and it is a very active  research area. In this paper, we are going to consider curvature based extended theories.

In particular, some models of ETG have been studied in the Newtonian limit \cite{PRD1,mio2}, as well as in the Minkowskian limit \cite{quadrupolo,Sta1,FOG_CGL2,FOGGW,leo,capriolo1,capriolo2}. The weak-field limit  has to be tested against realistic self-gravitating systems. Galactic rotation curves, stellar systems and gravitational lensing appear natural candidates as test-bed experiments~\cite{ BHL, BHL1, stabile_scelza,stabile_scelza2,stabstab,stabile_stabile_cap} (see also \cite{lv,Lambiase:2016bjy4,LambMohantySta}). In this perspective, corrections to GR were already considered by several authors
\cite{weyl_2,edd,lan,pauli,bach,buc,bic,FOG_CGL,FOG_CGL2,FOG_CGL3,FOG_CGL4,FOG_CGL5,FOG_CGL6,FOG_CGL7,FOG_CGL8,CasimirFOG,anu,tino,cqg,FOGST}.

Specifically, one may consider the generalization of $f(R)$ models, where $R$ is the Ricci scalar, through generic functions containing curvature invariants    such as Ricci squared terms  $R_{\alpha\beta}R^{\alpha\beta}$ or Riemann squared ones  $R_{\alpha\beta\gamma\delta}R^{\alpha\beta\gamma\delta}$.  These terms  are not independent  each other due to the Gauss-Bonnet topological invariant which establishes a relation among quadratic curvature invariants \cite{revelles,bogdanos}.

Due to the large amount of possible models, an important issue is to select viable ones by experiments and observations. In this perspective, the new born {\it multimessenger astronomy} is giving important  constraints to admit or exclude gravitational theories (see e.g. \cite{lombrisier,DeLaurentis}). However, also fine experiments can be conceived and realized in order to fix possible deviations and extensions with respect to GR. They can involve  space-based setups like satellites and precise electromagnetic measurements.

In this work we are going to analyze the exchange of  photons between the Earth and a satellite.  We model the Earth spacetime background by a spherical metric  that slowly rotates assuming a generic  ETG to model out the gravitational field. As a result, we find  a general relation for the photon frequency shift  to constrain the free parameters of the ETG  models, see also Ref. \cite{gaefre} and references therein.

The paper is organized as follows. In Sec.\ \ref{ETG1}, we summarize the weak field limit of   ETG models  considering a general theory where higher-order curvature invariants and a scalar field are included. In Sec. \ref{Frequency Shift in ETG},     the frequency shift of a photon, generated by a rotating gravitational source, is taken into account. In Sec. \ref{constraints}, we discuss the theoretical constraints on the considered ETG models. Finally, conclusions are drawn in Sec.\ \ref{conclusions}.


\section{Extended  Gravity}\label{ETG1}

A possible action for ETG is given by

\begin{eqnarray}\label{FOGaction}
\mathcal{S}\,=\,\int d^{4}x\sqrt{-g}\biggl[f(R,R_{\mu\nu}R^{\mu\nu},\phi)+\omega(\phi)\phi_{;\alpha}\phi^{;\alpha}+\mathcal{X}\mathcal{L}_m\biggr],
\end{eqnarray}
where $f$ is a generic function of the invariant $R$ (the Ricci scalar), the invariant $R_{\mu\nu}R^{\mu\nu}\,=\,Y$ ($R_{\mu\nu}$ is the Ricci tensor), the scalar field $\phi$, $g$ is the determinant of metric tensor $g_{\mu\nu}$ and\footnote{We use the convention $c\,=\,1$.} $\mathcal{X}\,=\,8\pi G$ is the standard gravitational coupling. The Lagrangian density $\mathcal{L}_m$ is the minimally coupled ordinary matter Lagrangian density, $\omega(\phi)$ is a generic function of the scalar field.

The field equations obtained by varying the action (\ref{FOGaction}) with respect to $g_{\mu\nu}$ and $\phi$, In the metric approach, are\footnote{We use, for the Ricci tensor, the convention
$R_{\mu\nu}={R^\sigma}_{\mu\sigma\nu}$, whilst for the Riemann
tensor we define ${R^\alpha}_{\beta\mu\nu}=\Gamma^\alpha_{\beta\nu,\mu}+\cdots$. The
affine connections are the  Christoffel symbols of the metric, namely
$\Gamma^\mu_{\alpha\beta}=\frac{1}{2}g^{\mu\sigma}(g_{\alpha\sigma,\beta}+g_{\beta\sigma,\alpha}
-g_{\alpha\beta,\sigma})$, and we adopt the signature is $(-,+,+,+)$.}:
\begin{eqnarray}
\label{fieldequationFOG}
&&f_RR_{\mu\nu}-\frac{f+\omega(\phi)\phi_{;\alpha}\phi^{;\alpha}}{2}g_{\mu\nu}-f_{R;\mu\nu}+g_{\mu\nu}\Box
f_R+2f_Y{R_\mu}^\alpha
R_{\alpha\nu}+
\\\nonumber\\\nonumber
 &&-2[f_Y{R^\alpha}_{(\mu}]_{;\nu)\alpha}+\Box[f_YR_{\mu\nu}]+[f_YR_{\alpha\beta}]^{;\alpha\beta}g_{\mu\nu}+\omega(\phi)\phi_{;\mu}\phi_{;\nu}\,=\,
\mathcal{X}\,T_{\mu\nu}\,,\nonumber\\
\nonumber\\
\label{FE_SF}
&&2\omega(\phi)\Box\phi+\omega_\phi(\phi)\phi_{;\alpha}\phi^{;\alpha}-f_\phi\,=\,0~.
\end{eqnarray}
where:
\[f_R\,=\,\frac{\partial f}{\partial R}, \,\,\,\,\,\,f_Y\,=\,\frac{\partial f}{\partial Y}, \,\,\,\,\,\,\omega_\phi\,=\,\frac{d\omega}{d\phi}\,, \,\,\,\,  f_\phi\,=\,\frac{d f}{d\phi}\,, \]
and $T_{\mu\nu}\,=\,-\frac{1}{\sqrt{-g}}\frac{\delta(\sqrt{-g}\mathcal{L}_m)}{\delta
g^{\mu\nu}}$ is the the energy-momentum tensor of matter.


Let us study the weak-field approximation and in the Newtonian limit of the theory. To do this, we perturb Eqs.~(\ref{fieldequationFOG}) and (\ref{FE_SF}) in a Minkowski background $\eta_{\mu\nu}$~\cite{PRD1}. We can set the perturbed expressions of metric tensor $g_{\mu\nu}$ and scalar field $\phi$ In the following way:
\begin{eqnarray}
\nonumber
&&g_{\mu\nu}\,\sim\,
\begin{pmatrix}
-1-g^{(2)}_{00}(t,\mathbf{x})-g^{(4)}_{00}(t,\mathbf{x})+\dots & g^{(3)}_{0i}(t,\mathbf{x})+\dots \\
g^{(3)}_{0i}(t,\mathbf{x})+\dots & +\delta_{ij}-g^{(2)}_{ij}(t,\mathbf{x})+\dots\end{pmatrix}\,=\,
\begin{pmatrix}
-1-2\Phi-2\Xi & 2A_i \\
\nonumber
2A_i & +\delta_{ij}-2\Psi\delta_{ij}\end{pmatrix}.
\end{eqnarray}
and
\begin{eqnarray}
\nonumber
&&\phi\,\sim\,\phi^{(0)}+\phi^{(2)}+\dots\,=\,\phi^{(0)}+\varphi.
\end{eqnarray}
We note that $\Phi$, $\Psi$, $\varphi$ are proportional to the power $c^{-2}$, $A_i$\footnote{$A_i$ are the components of a vector potential in the space coordinates $i=1,2,3$.} is proportional to $c^{-3}$  while $\Xi$ to $c^{-4}$. Introducing the following quantities:
\[{m_R}^2\,\doteq\,-\frac{f_R(0,0,\phi^{(0)})}{3f_{RR}(0,0,\phi^{(0)})+2f_Y(0,0,\phi^{(0)})}~,~~~~ {m_Y}^2\,\doteq\,\frac{f_R(0,0,\phi^{(0)})}{f_Y(0,0,\phi^{(0)})}~,~~~~{m_\phi}^2\,\doteq\,-\frac{f_{\phi\phi}(0,0,\phi^{(0)})}{2\omega(\phi^{(0)})}\,\]
the generic function $f$, up to  the $c^{-4}$ order,  can be developed as:
\begin{eqnarray}
\label{LimitFramework2}
f(R,R_{\alpha\beta}R^{\alpha\beta},\phi)\,=\,&&f_R(0,0,\phi^{(0)})\,R+\frac{f_{RR}(0,0,\phi^{(0)})}{2}\,R^2+\frac{f_{\phi\phi}(0,0,\phi^{(0)})}{2}(\phi-\phi^{(0)})^2\nonumber\\\\\nonumber&&+f_{R\phi}(0,0,\phi^{(0)})R\,\phi+f_Y(0,0,\phi^{(0)})R_{\alpha\beta}R^{\alpha\beta}~.
\end{eqnarray}
We note that the all other possible contributions in $f$ are negligible \cite{PRD1,mio2,FOG_CGL,FOGST}.

Considering matter as a perfect fluid, i.e. $T_{tt}\,=\,T^{(0)}_{tt}\,=\,\rho$ and $T_{ij}\,=\,T^{(0)}_{ij}\,=\,0$, following the references \cite{FOG_CGL,CasimirFOG} the gravitational potential $\Phi$, $\Psi$ and $A_i$ and the scalar field $\varphi$, for a ball-like source with radius $R$, take the form:
{\small
\begin{eqnarray}\label{ST_FOG_FE_NL_sol_ball}
\begin{array}{ll}
\Phi(\mathbf{x})\,=
\,-\frac{GM}{|\mathbf{x}|}\big[1+\zeta(|\mathbf{x}|)\big]~,
\\\\
\zeta(|\mathbf{x}|)\equiv g(\xi,\eta)\,F(m_+R)\,e^{-m_+|\mathbf{x}|}+\big[\frac{1}{3}-g(\xi,\eta)\big]\,F(m_-R)\,e^{-m_-|\mathbf{x}|}-
\frac{4\,F(m_YR)}{3}\,e^{-m_Y|\mathbf{x}|}\,,
\\\\
\Psi(\mathbf{x})\,=\,
-\frac{GM}{|\mathbf{x}|}\big[1-\psi(|\mathbf{x}|)\big]~,\\\\
\psi(|\mathbf{x}|)\,\equiv\,g(\xi,\eta)\,F(m_+R)\,e^{-m_+|\mathbf{x}|}+\big[\frac{1}{3}-g(\xi,\eta)\big]\,F(m_-R)\,e^{-m_-|\mathbf{x}|}+
\frac{2\,F(m_YR)}{3}\,e^{-m_Y|\mathbf{x}|}\,,
\\\\
\mathbf{A}(\mathbf{x})\,=\,
-\frac{2\,G\,\big[1-{\cal A}(|\mathbf{x}|)\big]}{|\mathbf{x}|^2}\,\mathbf{x}\times\mathbf{J}~,\\\\
{\cal A}(|\mathbf{x}|)\,\equiv\,(1+m_Y|\mathbf{x}|)\,e^{-m_Y|\mathbf{x}|}\,,
\\\\
\varphi(\textbf{x})\,=\,\sqrt{\frac{\xi}{3}}\,\frac{2 GM}{|\textbf{x}|}\biggl[\frac{F(m_+R)\,e^{-m_+\,|\textbf{x}|}\,-F(m_-R)\,e^{-m_-\,|\textbf{x}|}}{\omega_+-\omega_-}\biggr]~,
\end{array}
\end{eqnarray}}
where $\mathbf{J}\,=\,2M\mathcal{R}^2\mathbf{\Omega}_0/5$ is the angular momentum of the ball, $g(\xi,\eta)\,=\,\frac{1-\eta^2+\xi+\sqrt{\eta^4+(\xi-1)^2-2\eta^2(\xi+1)}}{6\sqrt{\eta^4+(\xi-1)^2-2\eta^2(\xi+1)}}\,$,

\[m_\pm^2=m_R^2\,\omega_\pm\,,\,\,\,\,\,\,\,\omega_\pm\,=\,\frac{1-\xi+\eta^2\pm\sqrt{(1-\xi+\eta^2)^2-4\eta^2}}{2}\,,\,\,\,\,\,\,\,\xi\,=\,3{f_{R\phi}(0,0,\phi^{(0)})}^2\,,\,\,\,\,\,\,\, \eta\,=\,\frac{m_\phi}{m_R}\] and $F(m\,\mathcal{R})\,=\,3\frac{m\,\mathcal{R} \cosh m\,\mathcal{R}-\sinh m\,\mathcal{R}}{m^3\mathcal{R}^3}$  \cite{FOGST},  $f_R(0,0,\phi^{(0)})\,=\,1$, $\omega(\phi^{(0)})\,=\,1/2$.  The metric can be written as:
\begin{eqnarray}\label{MeticCart}
&&g_{\mu\nu}\,\simeq\,
\begin{pmatrix}
-1-2\Phi & 2\textbf{A} \\
2\textbf{A} & (1-2\Psi)\delta_{ij}\end{pmatrix}~.
\end{eqnarray}
Starting from this result, we want to study the photon frequency shift in the framework of Extended Gravity.


\section{The Photon Frequency Shift in Extended Gravity}\label{Frequency Shift in ETG}

Let us consider a spherical planet of mass $M$ and angular momentum $J$  slowly rotating around itself. We suppose that the metric (\ref{MeticCart}) can be used to model the spacetime background around the rotating planet. For the sake of simplicity, we consider the equatorial plane, defined by $\theta=\pi /2$, where a photon is sent from an observer $A$ on Earth, at altitude $r=r_A$, to another observer $B$ on a satellite which sits on an orbit of radius  $r=r_B$.

In the equatorial plane, the metric (\ref{MeticCart}), in the coordinates $\{t,r,\theta,\tilde{\phi} \}$, reduces\footnote{Here, we are using the symbol $\tilde{\phi}$ to distinguish the coordinate with respet to the field $\phi$.} to:
\begin{eqnarray}\label{LineElement}
ds^2\,=\,- (1+2\Phi(r))dt^2+(1+2\Psi(r))(dr^2+r^2d\Omega^2)+2\,a(r)\,dtd \tilde{\phi}\,,
\end{eqnarray}
where $a(r)=|\textbf{A}(r)|=-2GJ{}{\cal A}/r$.

We denote with $\nu_X$ the frequency of the photon measured locally by the observer $X$, with $X\in\{A,\,B\}$ and the  proper time $\tau_X$. The observer $A$ on Earth prepares and sends a photon at altitude $r_A$ witch is received by the observer $B$ on the satellite at altitude $r_B$. The general frequency shift relation for a photon emitted from the observer $A$ and received by the observer $B$ reads \cite{SCH,Wald}:
\begin{eqnarray}\label{ShiftFreq}
F(r_A,r_B)=\frac{\nu_B}{\nu_A}=\frac{\biggl[ {\cal U}_\gamma^{\,\,\,\mu}\,{\cal U}_\mu{_{\,B}}\biggr]_{|\,r=r_B}}{\biggl[{\cal U}_\gamma^{\,\,\,\mu}\,{\cal U}_\mu{_{\,A}}\biggr]_{|\,r=r_A}}\,,
\end{eqnarray}
where ${\cal U}_\gamma^{\mu}\equiv(\dot{t}_\gamma,\,\dot{r}_\gamma,\,\dot{\theta}_\gamma,\,\dot{\tilde \phi}_\gamma)$ and  ${\cal U}_X^{\mu}\equiv(\dot{t}_X,\,\dot{r}_X,\,\dot{\theta}_X,\,\dot{\tilde \phi}_X)$ with $X\in\{A,\,B\}$ are the four-velocities of photon, observer $A$ and observer $B$, respectively; the dots stand for derivatives with respect to the proper time.

In our analysis, we consider the motion in the equatorial plane, $\theta=\pi/2$ and $\dot{\theta}_\gamma=\dot{\theta}_X=0$, and  the orbits of the two observer to be circular $\dot{r}_X=0$; moreover, we assume that the photon is sent radially, $\dot{\tilde \phi}_\gamma = 0$.

Therefore, we can write the scalar products in Eq. (\ref{ShiftFreq}) as:
\begin{eqnarray}\label{4Vel}
{\cal U}_\gamma^{\,\,\,\mu}\,{\cal U}_\mu{_{\,X}}=\dot{t}_\gamma\big(g_{tt}\dot{t}_X+g_{t\tilde\phi}\dot{\tilde\phi}_X\big)\,.
\end{eqnarray}
By making a Lagrangian analysis we can find the two conserved quantities for both observer and photon, i.e. the energy and the angular momentum as functions of $\dot{t}$ and $\dot{\tilde\phi}$, after some calculations we find:
\begin{eqnarray}\label{4Vel2}
\dot{t}&= &\,(1-2\Phi)\,E +\frac{a}{r^2}\,L\,,\\
\nonumber
\dot{\tilde\phi}&=&\,\frac{L}{r^2}  - \frac{a}{r^2}\,E\,.
\end{eqnarray}

Let us now determine the four-velocities of the photon and of the two observers $A$ and $B$. Since the photon is sent radially, its four-velocity can be written as ${\cal U}_\gamma^{\mu}=(\dot{t}_\gamma,\dot{r}_\gamma,0,0)$. From eq. (\ref{4Vel2}), we find that:
\begin{eqnarray}\label{4Vel23}
\dot{t}_\gamma &= &\,[1-2\Phi]\,E_\gamma\,\,\,\,\,\, \Rightarrow\,\,\,\,\,\, {\cal U}_\gamma^{\mu}=\big( (1-2\Phi)\,E_\gamma,\dot{r}_\gamma ,0,0 \big)\,.
\end{eqnarray}

The four-velocities of the two observers $A$ and $B$ are \cite{Chandra}:
\begin{eqnarray}\nonumber
{\cal U}_X^{\,\,\,\mu}\,=\Biggl[\frac{(1,0,0,\omega_X)}{\sqrt{1+2\Phi-(1-2\Psi)r^2\omega^2_X -2\,a\,\omega_X}}\Biggr]_{r=r_X}\text{with}\,\,\,\, X\in\{A,\,B\}\,.
\end{eqnarray}
For the observer $A$: the quantity $\omega_A=\dot{\tilde\phi}_A/\dot{t}_A$ is the angular velocity of the observer $A$ witch is not a geodesic, indeed it corresponds to source's equatorial angular velocity; For the observer $B$: the quantity $\omega_B=\dot{\tilde\phi}_B/\dot{t}_B$ is the angular velocity of the observer  $B$ on the satellite which follows a geodesic\footnote{This velocity can be expressed in terms of the Christoffel symbols.}. In both cases the normalization factor has been fixed by imposing the the condition: ${\cal U}_{\mu}\,{\cal U}^{\mu}=-1$.

Therefore, we now have all the ingredients to compute the frequencies $\nu_A$ and $\nu_B$ in Eq. (\ref{4Vel}), and so the corresponding shift. The frequencies measured by the observer $A$ and $B$ reads:
\begin{eqnarray}\nonumber
\biggl[{\cal U}_\gamma^{\,\,\,\mu}\,{\cal U}_\mu{_{\,X}}\biggr]_{r=r_X}\,=\,\Biggl[\frac{(1-2\Phi)E_\gamma[-(1+2\Phi)+a\,\omega_X]}{\sqrt{1+2\Phi-(1-2\Psi)r^2\omega^2_X -2\,a\,\omega_X}}\Biggr]_{r=r_X}\text{with}\,\,\,\, X\in\{A,\,B\}\,.
\end{eqnarray}
We can compute the quantity in eq (\ref{ShiftFreq}), which, up to linear order in the metric perturbation, is given by:
\begin{eqnarray}\label{ShiftFreq2}
F(r_A,r_B)=\frac{\nu_B}{\nu_A}\simeq&&\sqrt{1-r_A^2\omega^2_A}\biggl[ 1-\Phi(r_B)+\frac{r_B\Phi^{\prime}(r_B)}{2}+a(r_A)\omega_A\biggl]+\\
+&&\frac{1}{\sqrt{1-r_A^2\omega^2_A}}\biggl[ \Phi(r_A)-a(r_A)\omega_A+r^2_A\omega^2_A \Psi(r_A)\biggl]\,.
\end{eqnarray}
Note that the linearized frequency shift does not depend on whether the satellite orbit is direct or retrograde; it is also worth emphasizing that the presence of the square root implies $r^2_A\omega_A\leqslant1$, which is consistent with the fact that the tangential velocity on the Earth surface has to be smaller than the speed of light ($c = 1$).

The expression in eq. (\ref{ShiftFreq2}) can be further simplified by making a small angular velocity expansion, thus we obtain:
\begin{eqnarray}\label{ShiftFreq3}
F(r_A,r_B)&=& 1+\Phi(r_A)-\Phi(r_B)+\frac{r_B\Phi^{\prime}(r_B)}{2}+\\
&&-2\,r_A^2\omega^2_A\biggl[ 1-\Phi(r_A)-\Phi(r_B)+\frac{r_B\Phi^{\prime}(r_B)}{2}- 2 \Psi(r_A)\biggl]\,.
\end{eqnarray}

We now define the quantity $\delta$ \cite{gaefre,Fuentes}:
\begin{eqnarray}\label{Shift}
\delta(r_A,r_B)\,=1-\sqrt{F(r_A,r_B)}=\delta_{stat}(r_A,r_B)+\delta_{rot}(r_A,r_B)\,,
\end{eqnarray}
where
\begin{eqnarray}\label{Shift2}
\delta_{stat}(r_A,r_B)\,=-\frac{1}{2}\Biggl[ \Phi(r_A)-\Phi(r_B)+\frac{r_B\Phi^{\prime}(r_B)}{2}\Biggl] \,,
\end{eqnarray}
and
\begin{eqnarray}\label{Shift2rot}
\delta_{rot}(r_A,r_B)\,=\frac{\,r_A^2\omega^2_A}{4}\biggl[ 1-\frac{3}{2}\Phi(r_A)-\frac{\Phi(r_B)}{2}+\frac{r_B\Phi^{\prime}(r_B)}{2}- 2 \Psi(r_A)\biggl] \,,
\end{eqnarray}
through which we can quantify the frequency shift of a photon exchanged between the two observers $A$ and $B$; $\delta_{stat}$ and $\delta_{rot}$ quantify the static and rotation contribution, respectively.

We can immediately notice a very interesting result by working in the static case, $\omega_A=0$. Indeed, in the GR case, $\Phi=-G\,M/r$, there exist a special configuration $r_B = 3/2\, r_A$ for which
there is no frequency shift:
\begin{eqnarray}\label{ShiftStaticGR}
\delta ^{GR}_{stat}(r_A,r_B)\,=\frac{{\cal R}_S}{4\,r_A\,r_B}\Biggl[ r_B- \frac{3}{2}r_A\Biggr]\,,
\end{eqnarray}
where ${\cal R}_S$ is the source's Schwarzschild radius, ${\cal R}_S=2\,G\,M$. For $r_B = 3/2\, r_A$ we get $\delta_{stat}^{GR}(r_A,3/2\, r_A)=0$ which agrees with the result found in Ref. \cite{Fuentes}. For the Schwarzschild metric, this peculiar value of the distance corresponds to the location at which the gravitational shift induced by the Earth and the one induced by special relativity through the motion of the satellite, cancel each other \cite{gaefre,Fuentes}. For distances $r_B < 3/2\, r_A$, special relativity dominates so that the observer B sees the photon blue-shifted; while for $r_B > 3/2\, r_A$ the general relativistic effect becomes the dominant one and the photon is seen red-shifted by the observer B on the satellite.

\section{Experimental constraints}\label{constraints}

%

%
%

Our aim  is to test any possible compatibility of ETGs with  the experimental data. Let us consider the general relation for the change of frequency (\ref{Shift}) for a satellite on orbit with a generic radius $r_B$. Then, the relation (\ref{Shift}) can be put in the form:
\begin{eqnarray}\label{Shift3}
\delta(r_A,r_B)\,=\,\delta^{GR}_{stat}(r_A,r_B)+\delta^{GR}_{rot}(r_A,r_B)+\delta^{ETG}_{stat}(r_A,r_B)\,,
\end{eqnarray}
where
\begin{eqnarray}\label{Function2}
\delta^{GR}_{rot}(r_A,r_B)\,&=&\,\frac{(r_A\omega_A)^2}{4}\,,
\end{eqnarray}
and
\begin{eqnarray}\label{Function3}
\delta^{ETG}_{stat}(r_A,r_B)\,=\,[1+\zeta(r_A)\big]\delta^{GR}_{stat}(r_A,r_B)+\frac{3\,{\cal R}_S}{8\,r_A}\bigg[\zeta(r_A)-\frac{2}{3}\zeta(r_B)+\frac{2\,r_A}{3}\zeta^{\prime}(r_B)\bigg]\,.
\end{eqnarray}
If the satellite is located on the orbit with $r_B=3/2r_A$, we get:
\begin{eqnarray}\label{Function4}
\delta^{ETG}_{stat}(r_A,3/2r_A)\,=\frac{3\,{\cal R}_S}{8\,r_A}\bigg[\zeta(r_A)-\frac{2}{3}\zeta(3/2r_A)+\frac{2\,r_A}{3}\zeta^{\prime}(3/2r_A)\bigg]\,.
\end{eqnarray}
Introducing the following function $\Theta(m;r)$:
\[\Theta(m;r)=\,F(m\,r)\,e^{-m\,r}\bigg[ 1-\frac{2}{3}\big(1+m\,r\big)e^{-m\,r/2}\  \bigg]\,,\]
the eq. (\ref{Function4}) becames:
\begin{eqnarray}\label{Function5}
\delta^{ETG}_{stat}(r_A,3/2r_A)\,=\frac{3\,{\cal R}_S}{8\,r_A}\,\Lambda(m_+,m_-,m_Y,r_A)\,,
\end{eqnarray}
where
\begin{eqnarray}\label{Function6}
\Lambda(m_+,m_-,m_Y,r_A)= g(\xi,\eta)\Theta(m_+;r_A) +\bigg[\frac{1}{3}-g(\xi,\eta)\bigg]\Theta(m_-;r_A) -
\frac{4}{3}\Theta(m_Y;r_A)\,.
\end{eqnarray}
%
 %

Imposing the constraint $|\delta^{ETG}_{stat}|\lesssim \delta^{GR}_{rot}$, we obtain the relation:
\begin{eqnarray}
\label{Constraint}
\bigg| \Lambda(m_+,m_-,m_Y,r_A)\bigg|&\lesssim & \frac{2\,r_A^3\,\omega_A^2}{3\,{\cal R}_S}\,.
\end{eqnarray}

Here some examples of characteristic values\footnote{\textbf{Earth:} $M\simeq\, 5.97\times 10^{24}\,\text{Kg}$, $r_A\simeq\,2.6	\times10^{6}\text{m}$, $\omega_A\simeq\,7.27	\times10^{-5}\text{rad/s}$; \textbf{Moon:} $M\simeq\, 7.34\times 10^{22}\,\text{Kg}$, $r_A\simeq\,1.74	\times10^{6}\text{m}$, $\omega_A\simeq\,2.70	\times10^{-6}\text{rad/s}$; \textbf{Mars:}  $M\simeq\, 6.39\times 10^{23}\,\text{Kg}$, $r_A\simeq\,3.39	\times10^{6}\text{m}$, $\omega_A\simeq\,7.09	\times10^{-5}\text{rad/s}$; \textbf{Jupiter:} $M\simeq\, 1.89\times 10^{27}\,\text{Kg}$, $r_A\simeq\,69.91\times10^{6}\text{m}$, $\omega_A\simeq\,1.76	\times10^{-4}\text{rad/s}$.} for the relation (\ref{Constraint}), in the case of satellite on the orbit with $r_B=\frac{3}{2}r_A$:
\begin{eqnarray}\label{Constraint2}
 &&\Lambda \,\lesssim\,1.7\times 10^{-3} \,\,\,\,\,\,\,\,\,\,\,\text{Earth},\\
 \nonumber
 &&\Lambda \,\lesssim\,3.8\times 10^{-6} \,\,\,\,\,\,\,\,\,\,\,\text{Moon},\\
 \nonumber
 &&\Lambda \,\lesssim\,2.3\times 10^{-3} \,\,\,\,\,\,\,\,\,\,\,\text{Mars},\\
 \nonumber&&\Lambda \,\lesssim\,4.2\times 10^{-2} \,\,\,\,\,\,\,\,\,\,\,\text{Jupiter}\,.
\end{eqnarray}
Therefore, the best constraint is obtained by considering the Moon as gravitational source on which the observer $A$ is sitting. The angular velocity $\omega_A$ correspond with the Moon's angular velocity, $\omega_A=2.70\times 10^{-6}\text{rad/s}$ ; while the satellite, observer $B$, is in circular orbit around the Moon at distance $r_B=\frac{3}{2}\,r_A\simeq\,2.61	\times10^{6}\text{m}$. For this reason, the Moon is the best candidate to test ETG. In this case, we have:
\begin{eqnarray}\label{Constraint3}
| \Lambda(m_+,m_-,m_Y,r_A)|\lesssim 10^{-6}\,.
\end{eqnarray}

Summarizing, in GR there exists a peculiar satellite-distance $r_B=3/2\,r_A$ at which the static contribution to the frequency shift (\ref{ShiftStaticGR}) vanishes, since the effects induced by pure gravity and special relativity compensate. Then, on such a peculiar orbit in GR, we have the first non-vanishing contribution comes from the rotational term (\ref{Function2}). Such a property may not hold in ETG. In fact, we have the contribution (\ref{Function5}). Finally, the eq. (\ref{Constraint3}) imposes a constraint on the free parameters of ETG in agreement with the experimental data.

However, from another point of view, it is possible define an observation window in which any kind of detectable effect would imply the presence of new physics. As an example, let us consider the Moon as the gravitational source. In this case we have a static and rotational contributions given by $\delta_{stat}\sim \frac{1}{4} {\cal R}_S/r_A\simeq 1.57\times 10^{-11} $ and $\delta_{rot}\sim \frac{(r_A\,\omega_A)^2}{4} \simeq 1.38\times 10^{-16}$. Hence, we have the following observational window:
\begin{eqnarray}\label{Constraint36}
1.38\times 10^{-16}<\delta< 1.57\times 10^{-11}\,,
\end{eqnarray}
in which any kind of detectable effect do not depend on any GR effect, but would imply the presence of new physics. In the case of the Earth we have the following observation window \cite{gaefre} : $6.02\times 10^{-13}<\delta< 3.48\times 10^{-10}$. For the Moon the distance at which the static GR contribution to the shift vanishes is $r_B\simeq\,26071\text{km}$. Thus, if an experiment with a satellite in circular orbit at such a distance is performed, then any kind of non-vanishing detectable frequency shift falling in the range given by (\ref{Constraint36}), would represents an evidence of new physics.

\subsection*{Extended Gravity Models}

Let us now take into account some ETG models  studied in literature and reported in  Table \ref{table}.

{\begin{itemize}

\item Case A: $f(R)$ denotes a family of theories, each one defined by a different function $f$ of the Ricci scalar $R$. In this case the characteristic scale (mass) $m_R$ (see, Table \ref{table}, case A) depends only on the first and second derivatives of the function $f(R)$. Relation (\ref{Function6}) takes the form:
\begin{eqnarray}\nonumber
 \Lambda(m_R,\infty,\infty,r_A)\equiv\frac{1}{3}\Theta(m_R;r_A)\,,
\end{eqnarray}
then the constraint (\ref{Constraint3})  is:
\begin{eqnarray}
 \nonumber
 \frac{1}{3}F(m_R\,r_A)\,e^{-m_R\,r_A}\bigg[ 1-\frac{2}{3}\big(1+m_R\,r_A\big)e^{-m_R\,r_A/2}\bigg]\lesssim 10^{-6} \,.
\end{eqnarray}

 In literature, an interesting model of $f(R)$ gravity is  $f(R)\,=\,R-R^2/R_0$, where $R_0$ is a constant. This model is successfully adopted in cosmology \cite{staro}. In this case, the effective mass is $m_R^2=R_0/6$.

\item Case B:   $f(R,\,R_{\alpha\beta}R^{\alpha\beta})$, namely we also include the curvature invariant $R_{\alpha\beta}R^{\alpha\beta}$. In this case there are  two characteristic scales $m_R$ and $m_Y$ (see Table~\ref{table}). From (\ref{Function6}), becomes:
\begin{eqnarray}\nonumber
 \Lambda(m_R,\infty, m_Y,r_A)\equiv\frac{1}{3}\Theta(m_R;r_A)-\frac{4}{3}\Theta(m_Y;r_A)\,,
\end{eqnarray}
then the constraint (\ref{Constraint3})  is:
{\small \begin{eqnarray}\nonumber
\frac{1}{3}F(m_R\,r_A)\,e^{-m_R\,r_A}\bigg[ 1-\frac{2}{3}\big(1+m_R\,r_A\big)e^{-m_R\,r_A/2}\bigg]-\frac{4}{3}F(m_Y\,r_A)\,e^{-m_Y\,r_A}\bigg[ 1-\frac{2}{3}\big(1+m_Y\,r_A\big)e^{-m_Y\,r_A/2}\bigg]\lesssim 10^{-6}\,.
\end{eqnarray}}ù
As an illustration, we have $f(R,\,R_{\alpha\beta}R^{\alpha\beta})\,=\,R-R^2/R_0+R_{\alpha\beta}R^{\alpha\beta}/R_{ic}$, where $R_0$ and $R_{ic}$ are constants.

\item Case C: Scalar-tensor models $f(R,\,\phi)+\omega(\phi)\phi_{;\alpha}\phi^{;\alpha}$. There are two effective scales $m_+$ and $m_-$ generated from the interaction between gravity and the scalar field (see, Table \ref{table}, case C). The relation (\ref{Constraint3}), takes the form:
\begin{eqnarray}\nonumber
 \Lambda(m_+, m_-, \infty,r_A)\equiv g(\xi,\eta)\,\Theta(m_+;r_A)+\bigg[\frac{1}{3}-g(\xi,\eta)\bigg]\frac{4}{3}\,\Theta(m_-;r_A)\,,
\end{eqnarray}
then for the constraint (\ref{Constraint3}) we get:
{\small \begin{eqnarray}\nonumber
\frac{1}{3}\,g(\xi,\eta)\,(m_R\,r_A)\,e^{-m_R\,r_A}\bigg[ &&1-\frac{2}{3}\big(1+m_R\,r_A\big)e^{-m_R\,r_A/2}\bigg]+\\
\nonumber
&&+\bigg[\frac{1}{3}-g(\xi,\eta)\bigg]\,F(m_-\,r_A)\,e^{-m_-\,r_A}\bigg[ 1-\frac{2}{3}\big(1+m_-\,r_A\big)e^{-m_-\,r_A/2}\bigg]\lesssim 10^{-6}\,.
\end{eqnarray}}

As an particular example of scalar-tensor theory is:
\begin{eqnarray}\nonumber
f_{\rm ST}(R,\phi)\,=R-\frac{R^2}{R_0}+\lambda\,\varphi\,R+\frac{1}{2}\varphi_{,\alpha}\varphi^{,\alpha}-\frac{1}{2}{\cal M}^2\,\varphi^2~.
\end{eqnarray}
where $R_0$ ${\cal M}$ and $\lambda$ are constants.

\item Case D: $f(R,\,R_{\alpha\beta}R^{\alpha\beta},\phi)+\omega(\phi)\phi_{;\alpha}\phi^{;\alpha}$,  for which we have three effective scales $m_+$ and $m_-$ and $m_Y$. The relation (\ref{Constraint3}) in this case assume the expression:
\begin{eqnarray}\nonumber
 \Lambda(m_+,m_-, m_Y, r_A)\equiv g(\xi,\eta)\,\Theta(m_+;r_A)+\bigg[\frac{1}{3}-g(\xi,\eta)\bigg]\,\Theta(m_-;r_A)-\frac{4}{3}\Theta(m_Y;r_A)\,.
\end{eqnarray}
then for the constraint (\ref{Constraint3}) we get:
{\small \begin{eqnarray}\nonumber
&&\frac{1}{3}\,g(\xi,\eta)\,(m_R\,r_A)\,e^{-m_R\,r_A}\bigg[ 1-\frac{2}{3}\big(1+m_R\,r_A\big)e^{-m_R\,r_A/2}\bigg]+\\
\nonumber
&&\,\,\,\,\,\,\,\,\,\,\,\,\,\,\,\,\,\,\,\,\,\,\,\,\,\,\,\,\,\,\,\,\,\,\,\,\,\,\,\,\,\,\,\,\,\,\,\,\,\,\,\,\,\,\,\,\,\,\,\,\,\,\,\,\,\,+\bigg[\frac{1}{3}-g(\xi,\eta)\bigg]\,F(m_-\,r_A)\,e^{-m_-\,r_A}\bigg[ 1-\frac{2}{3}\big(1+m_-\,r_A\big)e^{-m_-\,r_A/2}\bigg]+\\
\nonumber
&&\,\,\,\,\,\,\,\,\,\,\,\,\,\,\,\,\,\,\,\,\,\,\,\,\,\,\,\,\,\,\,\,\,\,\,\,\,\,\,\,\,\,\,\,\,\,\,\,\,\,\,\,\,\,\,\,\,\,\,\,\,\,\,\,\,\,\,\,\,\,\,\,\,\,\,\,\,\,\,\,\,\,\,\,\,\,\,\,\,\,\,\,\,\,\,\,\,\,\,\,\,\,\,\,\,\,\,\,\,\,\,\,\,\,\,\,\,\,\,\,\,\,\,\,\,\,\,\,\,\,\,\,-\frac{4}{3}F(m_Y\,r_A)\,e^{-m_Y\,r_A}\bigg[ 1-\frac{2}{3}\big(1+m_Y\,r_A\big)e^{-m_Y\,r_A/2}\bigg]\lesssim 10^{-6}\,.
\end{eqnarray}}

 As a special case of   scalar-tensor-fourth-order gravity theory is the Non-Commutative Spectral Geometry (NCSG) \cite{connes_1,connes_2}. At a given cutoff scale (e.g. the Grand Unification scale), the  gravitational part of the action couples to the Higgs field ${\bf H}$ \cite{ccm} and the action reads
{\small \begin{eqnarray}\nonumber
{\cal S}_{\rm NCSG}\,=\,\int d^4x\sqrt{-g}\biggl[\frac{R}{2{\kappa_0}^2}+ \alpha_0\,C_{\mu\nu\rho\sigma}C^{\mu\nu\rho\sigma}
+\tau_0 R^\star R^\star+\frac{{\bf H}_{;\alpha}{\bf H}^{;\alpha}}{2}-{\mu_0}^2\,
{\bf H}^2-\frac{R\,{\bf H}^2}{12}+\lambda_0\,{\bf H}^4\biggr]\,,
\end{eqnarray}}
where $R^\star R^\star$ is the topological term related to the Euler characteristic, hence it is non-dynamical. Since the square of the Weyl tensor can be expressed in terms of $R^2$ and $R_{\mu\nu}R^{\mu\nu}$: $C_{\mu\nu\rho\sigma}C^{\mu\nu\rho\sigma}\,=\,2R_{\mu\nu}R^{\mu\nu}-\frac{2}{3}R^2$.

\begin{center}
\begin{table*}[ht]
{\small
\hfill{}
\begin{tabular}{|l|l|c|c|c|c|c|c|}
\hline
\multicolumn{1}{|c|}{\textbf{Case}}&\multicolumn{1}{|c|}{\textbf{ETG}}& \multicolumn{5}{|c|}{\textbf{Parameters}}\\
\cline{3-7}
& & $m^2_R$ & $m^2_Y$ &$m^2_\phi$&$m^2_+$&$m^2_-$
 \\
\hline
\hline
A&\tiny{$f(R)$ }&\tiny{$-\frac{f_{R}(0)}{3f_{RR}(0)}$}&$\infty$& 0 & $m^2_R$ & $\infty$ 
\\
\hline
\hline
B&\tiny{$f(R,R_{\alpha\beta}R^{\alpha\beta})$}&\tiny{$-\frac{f(0)}{3f_{RR}(0)+2f_Y(0)}$}&\tiny{$\frac{f_{R}(0)}{f_Y(0)}$}& 0 & $m^2_R$ & $\infty$
\\
\hline
\hline
C&\tiny{$f(R,\phi)+\omega(\phi)\phi_{;\alpha}\phi^{;\alpha}$}&\tiny{$-\frac{f_{R}(0)}{3f_{RR}(0)}$}& $\infty$ &\tiny{$-\frac{f_{\phi\phi}(0)}{2\omega(\phi^{(0)})}$}&\tiny{$m^2_R w_{+}$}&\tiny{$m^2_R w_{-}$}
\\
\hline
\hline
D&\tiny{$f(R,R_{\alpha\beta}R^{\alpha\beta},\phi)+\omega(\phi)\phi_{;\alpha}\phi^{;\alpha}$} &\tiny{$-\frac{f(0)}{3f_{RR}(0)+2f_Y(0)}$}&\tiny{$\frac{f_{R}(0)}{f_Y(0)}$}&\tiny{$-\frac{f_{\phi\phi}(0)}{2\omega(\phi^{(0)})}$}&\tiny{$m^2_R w_{+}$}&\tiny{$m^2_R w_{-}$}
\\
\hline
\end{tabular}}
\hfill{}
\caption{\label{table} We report  different cases of Extended Theories of Gravity including a scalar field and higher-order curvature terms. The free parameters  are given as effective masses with their  asymptotic behavior. Here, we assume  $f_R(0,\,0,\,\phi^{(0)})\,=\,1$,  $\omega(\phi^{(0)})\,=\,1/2$.}
\end{table*}
\end{center}

This case is particularly relevant because it represents a fundamental theory that, in principle, can be constrained by precision satellite experiments. See also \cite{FOG_CGL} for further details.

\end{itemize}}

\section{Conclusions}\label{conclusions}

In the context of ETGs, we  studied the frequency shift of  photons generated by rotating gravitational sources  in the weak field approximation. Specifically, we analyzed  photons traveling radially in the equatorial plane of  Earth, exchanged between an observer on  Earth and  another observer on a satellite in a circular orbit around  Earth. The aim is to  experimentally  constrain the free parameters of ETG models. These parameters can be  can be reduced to  effective masses or lengths as shown in Table I.

Specifically, we have seen that there is a particular circular orbit,  $r_B=\frac{3}{2}r_A$, with $A$ the position of the observer on the Earth and $B$  the  position of the observer on the satellite,  where the frequency of the received photons remains unchanged in a GR framework. In this case,     the system is  static and no rotation is assumed. This is an interesting result, because it provides a physical framework  where we can directly test  the possible frequency shifts due to ETG gravity contributions. However, in a realistic situation, we have to take in account also  rotational effects.

Finally, we obtained the general expression for the amount of  photon frequency  shift (\ref{Shift3}) in the context of ETGs. Assuming that  corrections induced on the photon frequency shift can be divided in a part due to the GR static field, another one to GR rotational field and another due to  ETG,    we find  relation  (\ref{Constraint}) to constrain the  free parameters of a given ETG. Finally, it is possible to  suggest  the Moon as a possible laboratory to set up this kind of experiments. In this perspective, the forthcoming space missions towards the Moon, Mars and other Solar System objects can be the arena for probing these effects  \cite{Moonlight}. A detailed discussion of these aspects will be developed in a forthcoming paper.

\section*{Acknowledgments}
The authors   acknowledge the support of  {\it Istituto Nazionale di Fisica Nucleare} (INFN) ({\it iniziative specifiche} MOONLIGHT2 and QGSKY).


\begin{thebibliography}{99}


\bibitem{riess}
         A.G. Riess  {\it et al.},
{\it The Astronomical Journal}  {\bf 116} (1998) 1009.

\bibitem{ast}
         S.  Perlmutter  {\it et al.},
{\it The Astrophysical Journal}  {\bf 517}  (1999) 565.

\bibitem{clo}
          S. Cole  {\it et al.},
 {\it Monthly Notices of the Royal Astronomical Society}  {\bf 362}, (2005) 505.

\bibitem{spe}
         S.D. Spergel  {\it et al.},
 {\it Astrophysical Journal Supplement Series}  {\bf 170}, (2007)  377.

\bibitem{carrol}
         S.M. Carroll, W.H. Press W.H., E.L. Turner,
 {\it Annual Review of Astronomy and Astrophysics}  {\bf 30}, (1992) 499.

\bibitem{sahini}
         V. Sahni, A.A. Starobinski,
{\it International Journal of Modern Physics D}  {\bf 9}, (2000) 373 .

\bibitem{Felix}
  S.~Capozziello and M.~De Laurentis,
  Phys.\ Rept.  {\bf 509} (2011) 167,

\bibitem{Nojiri}
  S.~Nojiri and S.~D.~Odintsov,
  Phys.\ Rept.\  {\bf 505} (2011) 59



\bibitem{vasilis}
  S.~Nojiri, S.~D.~Odintsov and V.~K.~Oikonomou,
  Phys.\ Rept.\  {\bf 692} (2017) 1

\bibitem{felice}
  A.~De Felice and S.~Tsujikawa,
  Living Rev.\ Rel.\  {\bf 13} (2010) 3

\bibitem{mag-fer-fra}
      G. Magnano, M. Ferraris,  M. Francaviglia,
 {\it General Relativity and Gravitation}  {\bf 19} (1987) 465.


\bibitem{ext1} G. J. Olmo, Int. J. Mod. Phys. D {\bf 20} (2011) 413.

\bibitem{ext2} N. Poplawski, Gen. Rel. Grav. {\bf 46} (2014) 1625.

\bibitem{ext3} Y. N. Obukhov, Int. J. Geom. Meth. Mod. Phys. {\bf 3},
 (2006) 95

\bibitem{ext4}
Y.-F. Cai, S. Capozziello, M. De Laurentis, and E. N. Saridakis, Rept. Prog. Phys. {\bf 79} (2016) 106901.

\bibitem{ext5} M. Hohmann, L. Ja\"rv, M. Krssak, and C. Pfeifer, PRD {\bf 97} (2018) 104042

\bibitem{ext6} A. Conroy,  and T. Koivisto, European Physical Journal C {\bf 78} (2018) 923.

\bibitem{PRD1}
   A.  Stabile,
{\it Physical Review D}  {\bf 82}, (2010) 064021.

\bibitem{mio2} A. Stabile, Phys. Rev. D {\bf 82}, (2010) 124026.

\bibitem{quadrupolo}
 M. De Laurentis and S. Capozziello,
{\it Astroparticle Physics}, {\bf 35} (2011) 257.

\bibitem{Sta1}
S.  Capozziello, and   A. Stabile, 
{\it Astrophys Space Sci} {\bf 358} (2015) 27

\bibitem{FOG_CGL2} G. Lambiase, M. Sakellariadou, A. Stabile, An. Stabile, JCAP {\bf 1507} (2015) 003.
\bibitem{FOGGW} G. Lambiase, M. Sakellariadou, A. Stabile, e-Print: 2012.00114.
\bibitem{leo} G. Lambiase, L. Mastrototaro, Astrophys.J. {\bf 904} (2020) 1, 19.

\bibitem{capriolo1}
  S.~Capozziello, M.~Capriolo and L.~Caso,
  Int.\ J.\ Geom.\ Meth.\ Mod.\ Phys.\  {\bf 16} (2019)  1950047

 \bibitem{capriolo2}
  S.~Capozziello, M.~Capriolo and S.~Nojiri,
  Phys.\ Lett.\ B {\bf 810} (2020) 135821



\bibitem{BHL}
         C.G.  Boehmer, T. Harko,  F.S.N. Lobo,
{\it Astroparticle Physics}  \textbf{29} (2008) 386.

\bibitem{BHL1}
        C.G.  Boehmer, T. Harko,  F.S.N. Lobo,
{\it Journal of Cosmology and Astroparticle Physics}  \textbf{0803} (2008) 024.

\bibitem{stabile_scelza}
          A. Stabile  and G. Scelza.,
{\it Physical Review D}  {\bf 84} (2012) 124023.

\bibitem{stabile_scelza2}
A. Stabile and G.  Scelza,
{\it Astrophys Space Sci} {\bf 357} (2015) 44.

\bibitem{stabile_stabile_cap}
         A.  Stabile, An. Stabile, S. Capozziello, {\it Physical Review D}  {\bf 88}  (2013) 124011.

\bibitem{stabstab} A. Stabile and An. Stabile, {\it Physical Review D}  {\bf 85} (2012) 044014. 



\bibitem{lv} M.~Blasone, G.~Lambiase, L.~Petruzziello and A.~Stabile, Eur.\ Phys.\ J.\ C {\bf 78} (2018)  976.


\bibitem{Lambiase:2016bjy4} L.~Buoninfante, G.~Lambiase, L.~Petruzziello and A.~Stabile, Eur.\ Phys.\ J.\ C {\bf 79} (2019)  41.

\bibitem{LambMohantySta} G. Lambiase,  S. Mohanty, A Stabile, Eur. Phys. J. C \textbf{78} (2018) 350.




\bibitem{weyl_2} H. Weyl, {\it Raum zeit Materie: Vorlesungen uuber allgemeine Relativitatstheorie},
Springer (1921) Berlin.

\bibitem{edd} A.S. Eddington,
\emph{The Mathematical Theory of Relativity},
Cambridge University Press (1924)  London.

\bibitem{lan} C.  Lanczos,
{\it Zeitschrift fur Physik A Hadrons and Nuclei,}   \textbf{73} (1932) 147.

\bibitem{pauli} W. Pauli,
{\it Phys. Zeit.}  \textbf{20} (1919) 457.

\bibitem{bach}
    R. Bach R.,
     {\it Mathematische Zeitschrift}  \textbf{9} (1921) 110.

\bibitem{buc}
 A.H.  Buchdahl,
{\it Il Nuovo Cimento}  \textbf{23} (1962) 141.

\bibitem{bic}
  G.V.  Bicknell,
{\it Journal of Physics A: Mathematical, Nuclear and General}  {\bf 7} (1974) 1061.

\bibitem{FOG_CGL}S. Capozziello, G. Lambiase, M. Sakellariadou, A. Stabile, An. Stabile, Phys.Rev. D {\bf 91} (2015) 044012.

\bibitem{FOG_CGL3}  G. Lambiase, M. Sakellariadou, A. Stabile, JCAP 1312 (2013) 020.

\bibitem{FOG_CGL4}  N. Radicella, G. Lambiase, L. Parisi, G. Vilasi, JCAP 1412 (2014) 014.

\bibitem{FOG_CGL5}  S. Capozziello, G. Lambiase, Int.J.Mod.Phys. D {\bf 12} (2003) 843.

\bibitem{FOG_CGL6}
  S. Calchi Novati, S. Capozziello, G. Lambiase, Grav. Cosmol. {\bf 6} (2000) 173.

\bibitem{FOG_CGL7}  S. Capozziello, G. Lambiase, H.J. Schmidt, Annalen Phys. {\bf 9} (2000)  39.

\bibitem{FOG_CGL8} S. Capozziello, G. Lambiase, G. Papini, G. Scarpetta, Phys. Lett. A {\bf 254} (1999) 11.

\bibitem{CasimirFOG} G. Lambiase, A. Stabile, An. Stabile, Phys.Rev. D {\bf 95} (2017)  084019.

\bibitem{anu} T. Biswas, E. Gerwick, T. Koivisto, A. Mazumdar, Phys. Rev. Lett. {\bf 108} (2012)  031101.
  L. Buoninfante, A.S. Koshelev, G. Lambiase, A. Mazumdar, JCAP 09 (2018) 034.
 L. Buoninfante, A.S. Koshelev, G. Lambiase, J. Marto, A. Mazumdar, JCAP 06 (2018) 014.
L. Buoninfante, G. Lambiase, A. Mazumdar, Nucl. Phys. B 944 (2019) 114646.

\bibitem{tino} G.M. Tino, L. Cacciapuoti, S. Capozziello, G. Lambiase, F. Sorrentino, Prog. Part. Nucl. Phys. {\bf 112} (2020) 103772

\bibitem{cqg} S. Capozziello, A. Stabile, Class. Quant. Grav. {\bf  26}, (2009) 085019.

\bibitem{FOGST} A. Stabile, S. Capozziello, Phys. Rev. D {\bf 87} (2013) 064002.



 \bibitem{revelles}
  M.~De Laurentis and A.~J.~Lopez-Revelles,
  Int.\ J.\ Geom.\ Meth.\ Mod.\ Phys.\  {\bf 11} (2014) 1450082

 \bibitem{bogdanos}
  C.~Bogdanos, S.~Capozziello, M.~De Laurentis and S.~Nesseris,
  Astropart.\ Phys.\  {\bf 34} (2010) 236

 \bibitem{lombrisier}
 L. Lombriser and A. Taylor,
 JCAP {\bf 03} (2016) 031.

\bibitem{DeLaurentis}
  M.~De Laurentis, O.~Porth, L.~Bovard, B.~Ahmedov and A.~Abdujabbarov,
  Phys.\ Rev.\ D {\bf 94} (2016)   124038


\bibitem{gaefre} L. Buoninfante, G. Lambiase, A. Stabile, Eur. Phys. J. C {\bf 80} (2020)  122



\bibitem{SCH} E. Schr\"{o}dinger, {\it Expanding universe}, Cambridge University Press (2011) Cambridge.

\bibitem{Wald} R. M. Wald, {\it General Relativity}, University of Chicago Press (1984) Chicago.

\bibitem{Chandra} S. Chandrasekhar, \textit{The Mathematical Theory of Black Holes} (Clarendon, Oxford, 1992)
\bibitem{Fuentes}  J. Kohlrus, D. E. Bruschi, J. Louko and I. Fuentes, EPJ Quant. Technol. 4 (2017) 7.
\bibitem{staro} Starobinsky, A.A., {\it Soviet Astronomy Letters}  \textbf{9} (1983)  302.
\bibitem{connes_1} A. Connes, {\sl Noncommutative Geometry}, Academic Press, New York (1994).
\bibitem{connes_2} A. Connes, M. Marcolli, {\sl Noncommutative Geometry, Quantum Fields and Motives}, Hindustan Book Agency,  India (2008)
\bibitem{ccm} A.H. Chamseddine, A. Connes,  and M. Marcolli,  {\it Adv. Theor. Math. Phys.}   {\bf 11} (2007) 991.


\bibitem{Moonlight}
 http://w3.lnf.infn.it/research/astroparticle-physics/moonlight-2/?lang=en.

\end{thebibliography}
\end{document}